\documentclass[twocolumn,showpacs,superscriptaddress,preprintnumbers,amsmath,amssymb,prc]{revtex4}

\usepackage{graphicx}
\usepackage{dcolumn}
\usepackage{bm}
\usepackage{float}
\usepackage{color}
\usepackage{CJK}
\usepackage[colorlinks,linkcolor=blue,urlcolor=blue,citecolor=blue]{hyperref}

\usepackage{isotope}
\usepackage[normalem]{ulem}
\renewcommand{\sout}{\bgroup \color{red} \ULdepth=-0.5ex \ULset}

\begin{document}
\begin{CJK*}{UTF8}{gbsn}

\title{Determining the temperature in heavy-ion collisions with multiplicity distribution}

\author{Yi-Dan Song}
\affiliation{Key Laboratory of Nuclear Physics and Ion-beam Application~(MOE), Institute of Modern Physics, Fudan University, Shanghai $200433$, China}

\author{Rui Wang}
\email{wangrui@sinap.ac.cn}
\affiliation{Shanghai Institute of Applied Physics, Chinese Academy of Sciences, Shanghai $201800$, China}

\author{Yu-Gang Ma}
\email{mayugang@fudan.edu.cn}
\affiliation{Key Laboratory of Nuclear Physics and Ion-beam Application~(MOE), Institute of Modern Physics, Fudan University, Shanghai $200433$, China}

\author{Xian-Gai Deng}
\affiliation{Key Laboratory of Nuclear Physics and Ion-beam Application~(MOE), Institute of Modern Physics, Fudan University, Shanghai $200433$, China}

\author{Huan-Ling Liu}
\affiliation{Shanghai Institute of Applied Physics, Chinese Academy of Sciences, Shanghai $201800$, China}
\date{\today}

\begin{abstract}
{By relating the charge multiplicity distribution and the temperature of a de-exciting nucleus through a deep neural network, we propose that the charge multiplicity distribution can be used as a thermometer of heavy-ion collisions.
Based on an isospin-dependent quantum molecular dynamics model, we study the caloric curve of reaction $\isotope[103]{Pd}$ $+$ $\isotope[9]{Be}$ with the apparent temperature determined through the charge multiplicity distribution.
The caloric curve shows a characteristic signature of nuclear liquid-gas phase transition around the apparent temperature $T_{\rm ap}$ $=$ $6.4~\rm MeV$, which is consistent with that through a traditional heavy-ion collision thermometer, and indicates the viability of determining the temperature in heavy-ion collisions with multiplicity distribution.}

\end{abstract}
\maketitle

\section{Introduction}\label{1}

Understanding the properties of nuclear matter is one of the major objectives in nuclear physics.
At zero temperature, the properties of nuclear matter have been studied extensively, and its equation of state~(EOS), including its isospin dependence, i.e., symmetry energy, has been determined relatively well~\cite{LatPR442,BalPPNP91,RocPPNP101}, while its properties at finite temperature are relatively little touched upon.
Among these properties, two noticeable examples are the nuclear liquid-gas phase transition~\cite{FinPRL49,SieNt305,PanPRL52,PocPRL75,MYGPLB390,MYGPRL83,ChoPRL85,RicPR350,NatPRC65,NatPRL89,MYGPRC71,MCWPPNP99,BorPPNP105} and the temperature dependence of the ratio of shear viscosity to entropy density~($\eta/s$)~\cite{AuePRL103,DanPRC84,FDQPRC89,DXGPRC94,MonPRL118,GCQPRC95}.
The latter is also connected to the nuclear giant dipole resonance at finite temperature~\cite{BraPRL62,BorPRL67,WiePRL97}, 
since both of them are related to the two-body dissipation of nucleons. 

The difficulties of studying the finite temperature properties of nuclear matter mainly come from the preparation of a finite temperature nuclear system, as well as the determination of its temperature.
Heavy-ion collisions~(HICs) at intermediate-to-low energies provide a possible venue of investigating the finite temperature properties of nuclear matter~\cite{WadPRC39}.
During the reaction, a transient excited system is formed, and commonly it can be regarded as a (near)-equilibrium state~\cite{dEnPRL87,BorPLB388}, since the evolution of its constituent nucleons is sufficiently short comparing with the global evolution.
Its temperature can be accessed by, e.g., energy spectra through moving source fitting~\cite{WadPRC39}, excited state populations~\cite{ChiPLB172,SchPRC48}, (double)-isotope ratios~\cite{TsaPRL78,SerPRL80,AlbILNCA89}, or quadruple momentum fluctuation~\cite{WueNPA843} etc.
For a reliable thermometer of HICs, we require it is insensitive to both the collective effects and the secondary decay of unstable nuclei after the system disintegrates, which is commonly hard to achieve.
Besides that, because of the difficulty of examining the accuracy of the apparent temperature obtained through these thermometers, it is  not a trivial task to propose different ways of determining the apparent temperature, and thus provide more opportunities of crosscheck.

Machine-learning techniques~\cite{LeCNt521,JorSc349}, which have been applied extensively in physics~\cite{CarRMP91} due to their ability of recognizing and characterizing complex sets of data, provide an alternative and peculiar way of determining the apparent temperature of HICs at intermediate-to-low energies.
Besides the common uses like particle identification and tagging in experiments, machine-learning techniques have various novel applications in physics, e.g., solving quantum many-body problem~\cite{CarSc355,ZZWCPL37}, analyzing strong gravitational lenses~\cite{HezNt548}, classifying phases of matter~\cite{CarNtP13,NieNtP13,RodNtP15,RWJCPL37}, extrapolating the cross section of nuclear reactions~\cite{MCWCPC44}, and instructing single crystal growth~\cite{YTSCPL36} and optimizing experimental control \cite{WYCPL37}.
In a recent work of studying nuclear liquid-gas phase transition with machine-learning techniques~\cite{WRPRR2}, it has been shown that machine-learning techniques can capture the essential features of HICs directly from experimental final-state charge multiplicity distribution.

In the present letter, we propose that, with the help of machine-learning technique, the charge multiplicity distribution can be applied to determine the apparent temperature of HICs at intermediate-to-low energies.
Our basic methodology is as follows.
We first obtain the multiplicity of fragments of an excited nuclear source with given temperature based on a theoretical model, e.g., transport models~\cite{BerPR160,AicPR202}, statistical models~\cite{BonPR257,ChaNPA483} or their hybrid~\cite{GaiPPNP62,ZZFNST29,OnoPPNP105}.
We then relate the final-state charge multiplicity distribution with the temperature of the source via a deep neural network~(DNN).
This relation can be employed to determine the apparent temperature of a certain transient state during HICs at intermediate-to-low energies through their final-state charge multiplicity distribution.
We use the above method to determine the apparent temperature of a fragmentation reaction $\isotope[103]{Pd}$ $+$ $\isotope[9]{Be}$.
We further test the viability of the present method by comparing it with the momentum fluctuation thermometer~\cite{WueNPA843}, and by studying the characteristic signature of the caloric curve with the apparent temperature determined through the above method.
To adopt the above method to realistic HICs analyses with experimental charge multiplicity distribution relies on the precise determination of the fragment multiplicities from theoretical model.
Nevertheless, as a viability quest, in the present work, we employ an isospin-dependent quantum molecular dynamics (IQMD) model~\cite{MYGPRC73} to simulate the de-excitation of the finite temperature nuclear source, and do not require it to describe precisely the experimental fragment multiplicities.
More accurate description of the experimental fragment multiplicities can be achieved through, e.g., combining certain transport model with statistical model of multi-fragmentation~\cite{OnoPPNP105}, and it is beyond the scope of the present work.

\section{Methodology}\label{2}

\begin{figure}[htbp]
\includegraphics
[width=7.5cm]{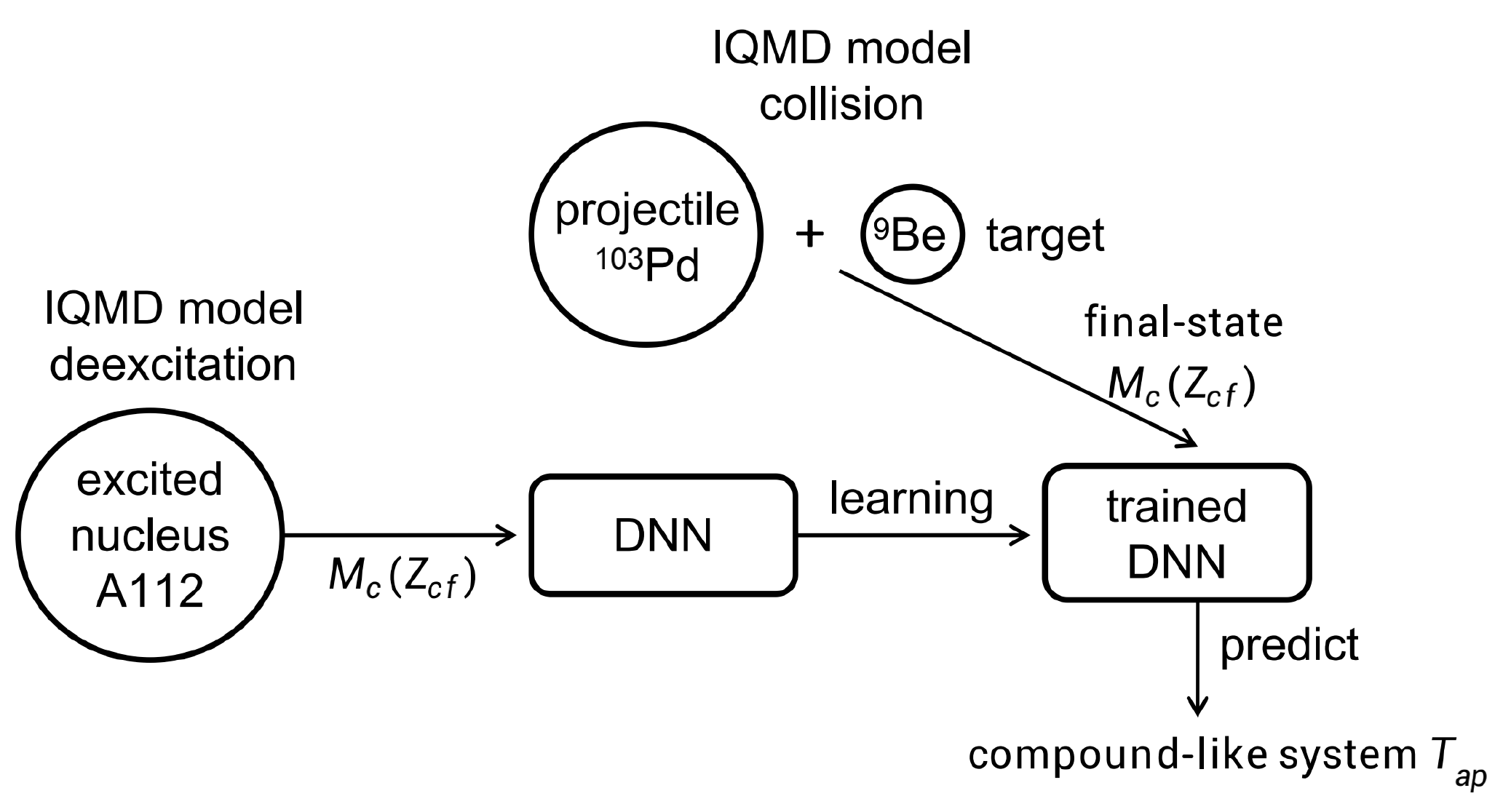}
\caption{The basic procedure of determining the $T_{ap}$ of $\isotope[103]{Pd}$ $+$ $\isotope[9]{Be}$ through its final-state $M_{\rm c}(Z_{\rm cf})$ with DNN.} \label{fig.1}
\end{figure}

In the present work, we focus on fragmentation reactions~\cite{SumPRC61,SYDNST29}, i.e., central $\isotope[103]{Pd}$ $+$ $\isotope[9]{Be}$ collision with incident energies ranging from 20 to 400$A$ MeV, and try to determine their apparent temperature through its final-state charge multiplicity distribution $M_{\rm c}(Z_{\rm cf})$, where $Z_{\rm cf}$ represents the charge number of the final charged fragments, ranging from $1$ to the charge number of the reaction system.
In the process of intermediate-to-low energy HICs, especially for fragmentation reactions, the two incident nuclei collide and then form a compound-like system.
This compound-like system is regarded as a (near-)equilibrium system, and we can mimic approximately this transient state of the reaction by a nuclear source or a finite nucleus at given temperature, with mass number $A$ and proton number $Z$, which are the same as those in the reaction $\isotope[103]{Pd}$ $+$ $\isotope[9]{Be}$~\cite{ChaPRC82}.
This finite temperature nuclear source has been employed to study finite-size scaling phenomenon~\cite{LHLPRC99}.
We first simulate the evolution of the nuclear source $^{112}$Sn ($A = 112$ and $Z = 50$) at different initial temperature $T^{model}$, and obtain their $M_{\rm c}(Z_{\rm cf})$ based on the IQMD model.
These simulations provide us $M_{\rm c}(Z_{\rm cf})$ from a nuclear source with a given temperature, which is difficult to obtain directly from HICs.
We then establish a relation between the source temperature $T^{model}$ and $M_{\rm c}(Z_{\rm cf})$ through a DNN, which recasts the complex relation into a non-linear map through its neurons.
Based on this relation, the apparent temperature of the reaction $\isotope[103]{Pd}$ $+$ $\isotope[9]{Be}$ can be determined through its final-state $M_{\rm c}(Z_{\rm cf})$.
In Fig.~\ref{fig.1}, we show the basic procedure of the proposed method to determine apparent temperature of  fragmentation reaction $\isotope[103]{Pd}$ $+$ $\isotope[9]{Be}$ through the nuclear source $^{112}$Sn.

\subsection{IQMD model}
The IQMD model, a many-body theory to describe the dynamics of HICs, can be derived from a time-dependent Hartree theory with Gaussian single-particle wave function $\phi_{i}(\vec{r},t)$,
\begin{equation}\label{eq.1}
\phi_{i}(\vec{r},t) =
\frac{1}{(2{\pi}L)^{3/4}}{\rm exp}\Big{[}-\frac{[\vec{r}-\vec{r}_{i}(t)]^2}{4L}+\frac{i\vec{p}_{i}(t)\cdot\vec{r}}{\hbar}\Big{]},
\end{equation}
with its spatial center $\vec{r}_{i}(t)$ and momentum center $\vec{p}_{i}(t)$ as variational parameters.
Other quantities can be obtained subsequently through $\vec{r}_{i}(t)$ and $\vec{p}_{i}(t)$.
In the above equation, $L$ is the square of the width of Gaussian wave packet and is set to be 2.18 $\rm fm^{2}$.
Through a product of single-particle wave functions $\phi_{i}(\vec{r},t)$, we can get the system wave function,
\begin{equation}\label{eq.6}
\psi(\vec{r}_{1},...\vec{r}_{n},t) = \prod_{i=1}^{A}\phi_{i}(\vec{r},t),
\end{equation}
where $A$ is the mass number of the system.
The potential energy $U$ in the IQMD model includes Skyrme, Yukawa, symmetry, momentum-dependent and Coulomb terms,
\begin{equation}\label{eq.2}
U = U_\text{Sky} + U_\text{Yuk} + U_\text{sym} + U_\text{MDI} + U_\text{Coul}.
\end{equation}
Detailed descriptions of the IQMD model, including the potential and equations of motion of $\vec{r}_{i}(t)$ and $\vec{p}_{i}(t)$, can be found in Ref.~\cite{HarEPJA1,MYGPRC73}. Numerous applications of IQMD have been made for different observable, and some recent ones can be found in Ref.~\cite{LPCNST29,FZQNST29,YTZNST30a,YTZNST30b,YHNST31}.

In the IQMD model, the initial nucleons are generated through the local density approximation.
For a ground-state nucleus, we generate $A$ Gaussian wave-packets, with their spatial center $\vec{r}_i$ $(t=0)$ sampled randomly within a sphere of radius given by $r_0A^{1/3}$ with $r_0$ $=$ $1.12~\rm fm$, and their momentum center $\vec{p}_i$ $(t=0)$ sampled following the zero-temperature Fermi-Dirac distribution.
A finite nucleus at a given temperature $T$ can be generated by sampling $\vec{p}_i$ $(t=0)$ according to finite temperature Fermi-Dirac distribution \cite{FDQPRC89}, which is given by,
\begin{equation}
\begin{split}
f(\epsilon)= \frac{1}{\exp(\frac{\epsilon-\mu}{T})+1},
\end{split}
\label{FermiDiracDis}
\end{equation}
where $\epsilon$ $=$ $\sqrt{p^{2}+m^{2}}$ $+$ $U$ is the single particle energy, and $\mu$ is the chemical potential.
The parameters $p$, $m$ and $U$ are momentum, mass and potential energy, respectively.
For simplicity, we have omitted the contribution from momentum-dependent part in the above equation.

\subsection{Deep neural network}
\begin{figure}[htbp]
\includegraphics
[width=7.5cm]{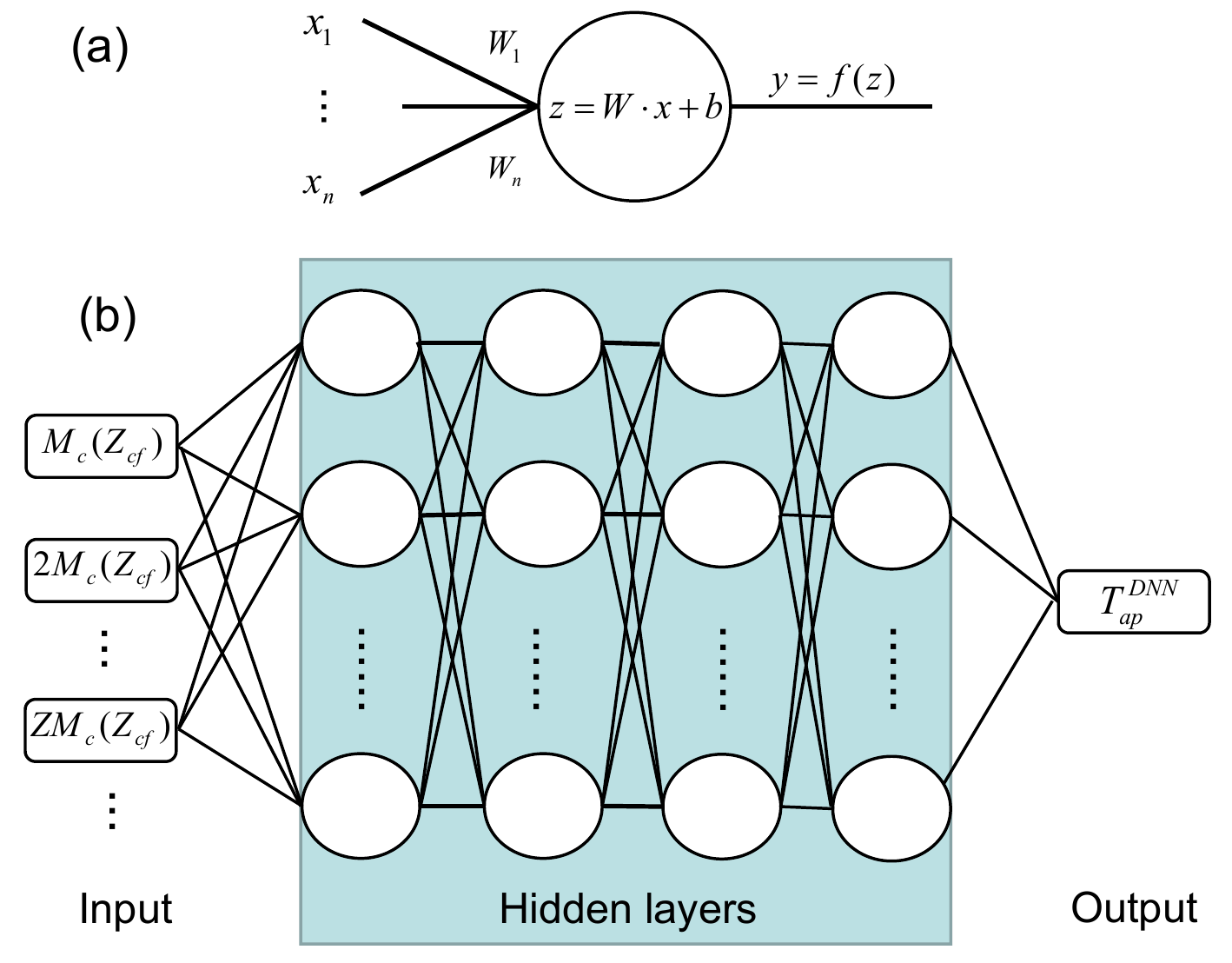}
\caption{(a). A single artificial neuron, with $n$ inputs labelled as $x_{1}$ through $x_{n}$ and an output $y$.
The output of the neuron is computed by applying the activation function $f(z)$ to the product of $\textbf{x}$, weights $W$, and biases $b$, e.g., $\textbf{z}$ = $\textbf{W}\cdot\textbf{x}$ $+$ $\textbf{b}$.
(b). The feed-forward neural network used in the present work, consisting of an input layer, an output layer and four hidden layers. The input of the network is the charge-weighted charge multiplicity distribution $ZM_c(Z_{cf})$.} \label{fig.6}
\end{figure}

In the present work, we adopt a feed-forward DNN to establish the relation between the temperature $T^{\rm Model}$ of the de-exciting nuclear source generated in IQMD and its final-state charge multiplicity distribution $M_{\rm c}(Z_{\rm cf})$.
We treat $M_{\rm c}(Z_{\rm cf})$ as the input image, while its corresponding temperature as its label.
The DNN contains successively one input layer, several hidden layers, and one output layer.
Each layer generates its output $\textbf{z}$ through a matrix multiplication of its input $\textbf{x}$, i.e., $\textbf{z}$ = $\textbf{W}\cdot\textbf{x}$ $+$ $\textbf{b}$.
The elements in the matrix $\textbf{W}$ are known as weights and in the vector $\textbf{b}$ as biases.
In a normal full-connected neural network these parameters are single values.
The neuron is then followed by an activation function $f(\textbf{z})$, which turns a linear transform to a non-linear one.
Commonly used activation functions are \emph{sigmoid}, \emph{tanh}, and \emph{ReLU}~(rectified linear unit).
$f(\textbf{z})$ is then used as the input of the next layer.
In the present work, the neural network can be treated as a functional
\begin{equation}\label{E:DNN}
    T_{\rm ap}^{\rm DNN} = g\big[M_{\rm c}(Z_{\rm cf});\textbf{W},\textbf{b}\big],
\end{equation}
which relates non-linearly a given input $M_{\rm c}(Z_{\rm cf})$, a vector with $50$ elements in our case, to an output predicted apparent temperature $T_{\rm ap}^{\rm DNN}$.
We employ four hidden layers, with each consists of 32 artificial neurons.
The input layer and the first three hidden layers are followed by \emph{ReLU}, while the last hidden layer and the output predicted $T_{\rm ap}^{\rm DNN}$ are connected directly by the matrix multiplication.
The sketch of the DNN used in the present work can be seen in Fig.~\ref{fig.6}.    

We train the network based on a data set $\big\{M_{\rm c}(Z_{\rm cf}),T^{\rm Model}\big\}$, to minimize the cost function, the difference between the given temperature and the DNN prediction, i.e., $(T^{\rm model} - T_{\rm ap}^{\rm DNN})^2$, by adjusting its parameters $\textbf{W}$ and $\textbf{b}$.
The optimization is fulfilled by the $Adam$~\cite{Adam} package in \emph{Tensorflow}.
During training the network, we use an exponential decreasing learning rate $\alpha$ $=$ $10^{-6}$ $+$ $(10^{-3} - 10^{-6})\exp(-i/10000)$, with $i$ the training epoch and it is equal to 10,000.
To prevent the network from over fitting the data set, we include a standard $l_2$ regularization term, i.e., a term proportional to the norm of the weight $\textbf{W}$ and the bias $\textbf{b}$, $l_2(\Vert\textbf{W}\Vert^2/2 + \Vert\textbf{b}\Vert^2/2)$, with $l_2$ a positive number, in the cost function of the neural network.
The $l_2$ regularization prevents the weights and biases from increasing to arbitrary large values during the optimization.

\section{Results and discussion} \label{3}

\subsection{Apparent temperature}
\begin{figure}[htbp]
\includegraphics
[width=7.5cm]{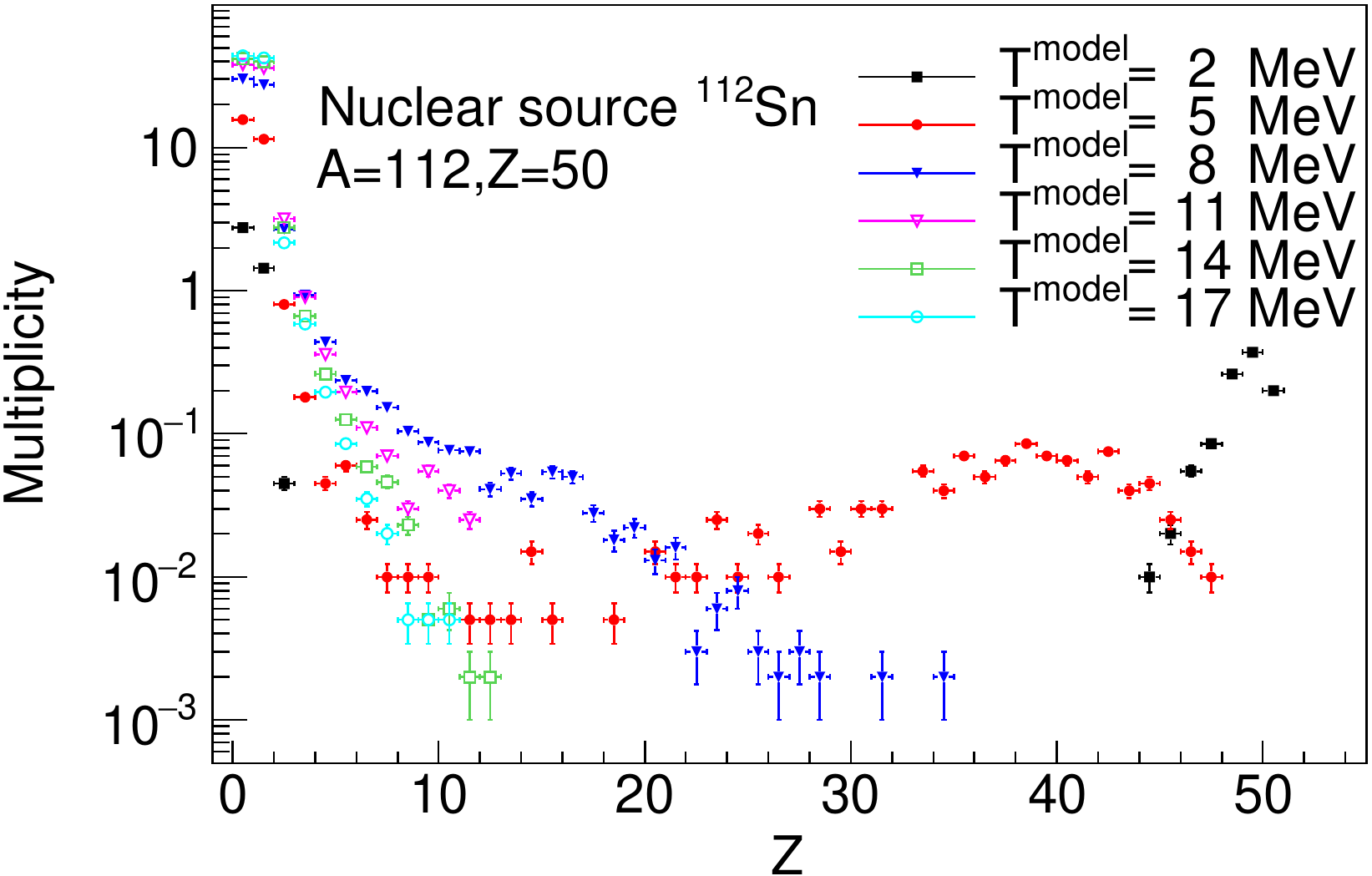}
\caption{The charge distribution of fragments from excited nuclear source $\isotope[112]{Sn}$ at several $T^{\rm model}$. 
For each $T^{\rm model}$, we show the simulated data of one run consisting of 2,000 events from the IQMD model.} \label{fig.10}
\end{figure}

Based on the IQMD model, we simulate the de-excitation process of a $\isotope[112]{Sn}$ nucleus at different $T^{\rm model}$.
The \isotope[112]{Sn} nucleus has the same mass and charge number with the $\isotope[103]{Pd}$ $+$ $\isotope[9]{Be}$ reaction, and it can be regarded as a nuclear source to mimic the transient excited state of the reaction.
We perform $50$ runs for each $T^{\rm model}$, with each run consists of $2,000$ events.
$T^{\rm model}$ ranges from $0$ to $20~\rm MeV$ with $1~\rm MeV$ interval.
Fig.~\ref{fig.10} displays charge distributions of fragments from hot nuclear source \isotope[112]{Sn} at several $T^{\rm model}$ with the data sample of one run~($2,000$ events) at each temperature.
These fragment charge distributions exhibit typical changes of disassembly mechanism of hot nuclei with temperature \cite{MYGPRC71,MYGPRL83}, i.e., from evaporation mechanism at lower temperature (e.g.  $T^{\rm model}$= 2 MeV), to multifragmentation at medium temperature (e.g.  $T^{\rm model}$= 8 MeV), till vaporization at higher temperature (e.g.  $T^{\rm model}$= 17 MeV). According to the shapes of these charge distributions versus temperature, a nuclear liquid-gas phase transition shall begin to happen at a certain moderate temperature for this system \cite{MYGPRC71,LHLPRC99,WRPRR2,WadPRC99}.

\begin{figure}[htbp]
\includegraphics
[width=7.5cm]{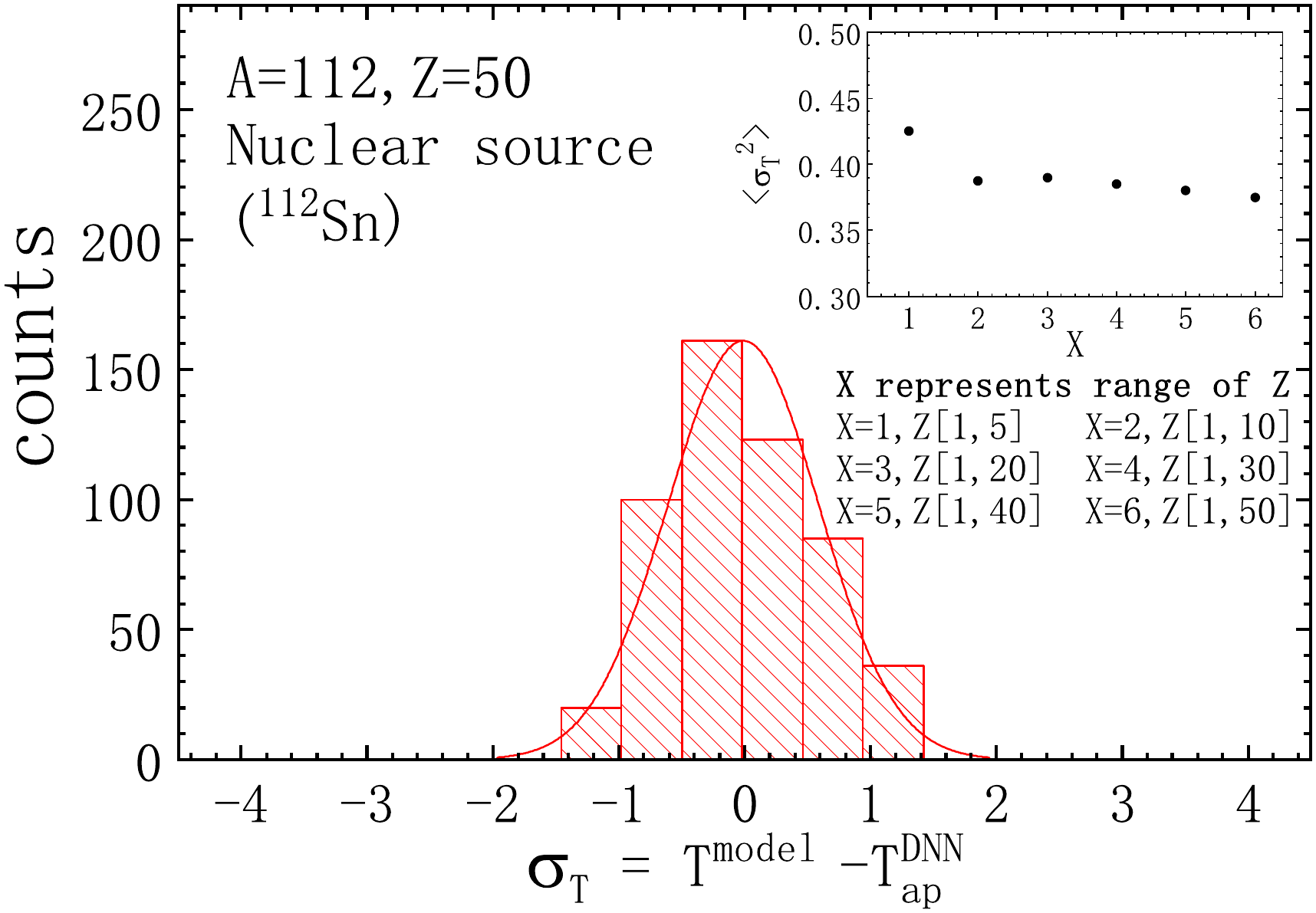}
\caption{The histogram of errors $\sigma_{\rm T}$ between the predicted apparent temperature from DNN $T_{\rm ap}^{\rm DNN}$ and the original $T^{\rm model}$ given in the IQMD simulations for the excited nuclear source (finite temperature \isotope[112]{Sn}).
The red line is the Gaussian fitting of the histogram.
The inset shows the testing accuracy $\langle\sigma_{\rm T}^{2}\rangle$ for different charge multiplicity distributions.
$X$ represents different ranges of charge multiplicity distribution.}
\label{fig.2}
\end{figure}

We can train the DNN once we get the above-mentioned charge multiplicity distributions $M_{\rm c}(Z_{\rm cf})$. 
In the present work, the total $50\times21$ charge multiplicity distributions $M_{\rm c}(Z_{\rm cf})$, and their corresponding $T^{\rm model}$, are treated as the images and labels, respectively, and they form the data set $\big\{M_{\rm c}(Z_{\rm cf}),T^{\rm model}\big\}$ of the DNN.
The data set is further divided into training set and testing set, each contains half of the total data set.
After we trained the DNN with the training set, i.e., determining the parameters \textbf{W} and \textbf{b} in Eq.~(\ref{E:DNN}) to best reproduce the given $T^{\rm model}$, one can predict the apparent temperature with a given $M_{\rm c}(Z_{\rm cf})$ through Eq.~(\ref{E:DNN}).
We show in Fig.\ref{fig.2} the histogram of the DNN's predicting error $\sigma_{\rm T}$, i.e., the difference between the original $T^{\rm model}$ and its DNN prediction $T_{\rm ap}^{\rm DNN}$, of the testing set.
The standard error of the DNN prediction is about $0.62~\rm MeV$, which is small enough for further analyses based on the apparent temperature obtained through the present way.

In the inset of Fig.~\ref{fig.2}, we show the dependence of the testing accuracy on the range of charge multiplicity distribution, i.e., only the charge multiplicity distributions within the given range~(represented by $X$ in the inset) are used in training and testing the DNN.
We notice from the inset that the light fragments, i.e., $Z\in[1,5]$ play major role in determining the apparent temperature, while including heavier fragments do help to increase the accuracy.
In another perspective, this feature actually indicates the superiority of the present method to the traditional isotopic ratio thermometer, since for the later only the information of light fragments is taken into consideration. 
At present, we only use the charge multiplicity distribution to predict the apparent temperature by the DNN. 
In principle, the information in momentum space can be included in the present method for better accuracy.

\begin{figure}[htbp]
\includegraphics
[width=7.5cm]{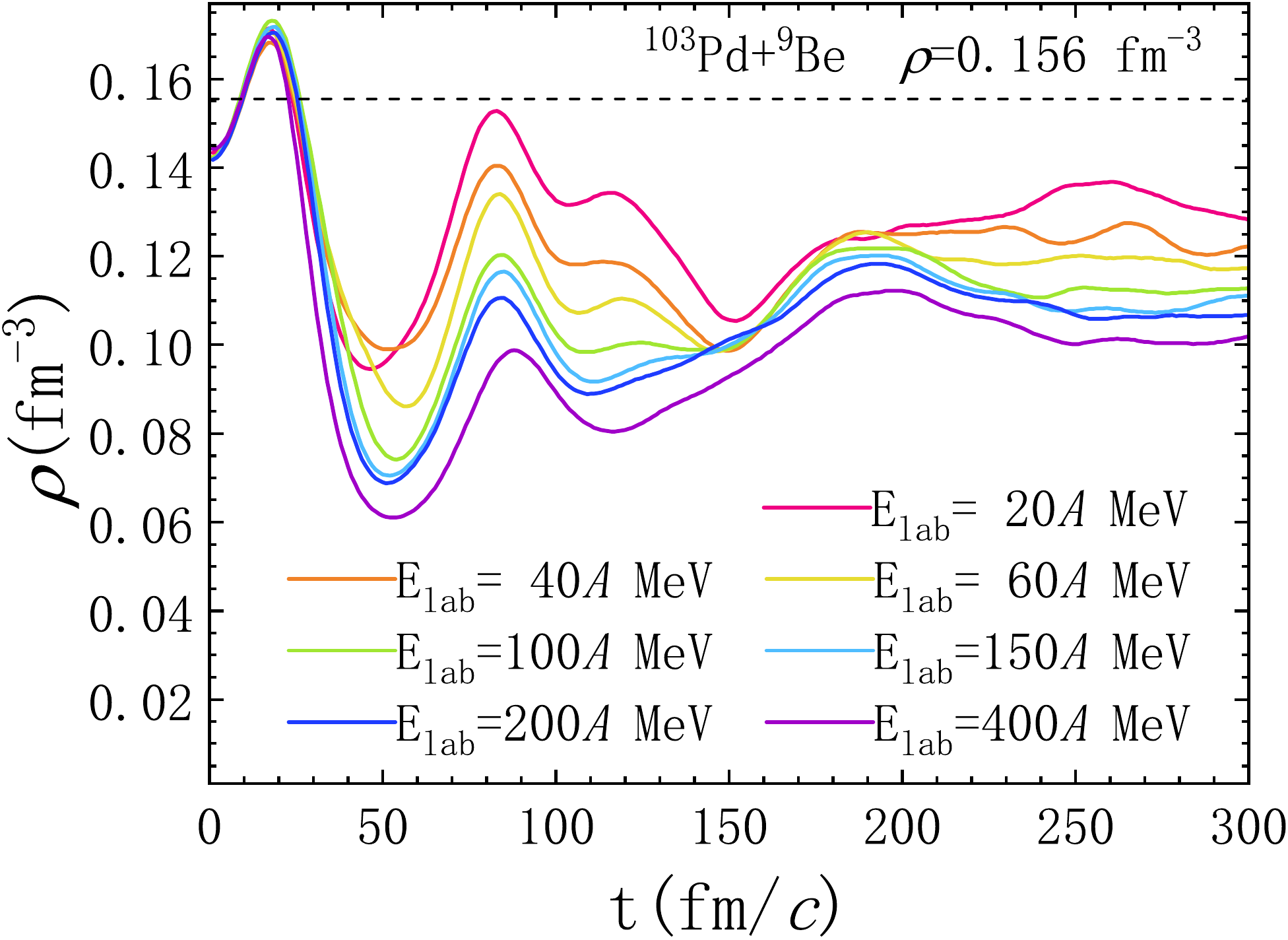}
\caption{Time evolution of the central density of the heaviest fragment in reaction $\isotope[103]{Pd}$ $+$ $\isotope[9]{Be}$ with the IQMD simulation.
Different lines denote the results from different incident energies from 20$A$ MeV to 400$A$ MeV.
The black dashed line represents the initial central density of the nuclear source \isotope[112]{Sn}.} \label{fig.3}
\end{figure}

After establishing the relation between the apparent temperature and $M_{\rm c}(Z_{\rm cf})$ through training the DNN, we turn to determine the apparent temperature of $\isotope[103]{Pd}$ $+$ $\isotope[9]{Be}$ fragmentation reactions.
We first examine the reaction dynamics of $\isotope[103]{Pd}$ $+$ $\isotope[9]{Be}$ within the IQMD model.
We simulate the reactions $\isotope[103]{Pd}$ $+$ $\isotope[9]{Be}$ with incident energy $E_{\rm lab}$ ranging from 20$A$ MeV to 400$A$ MeV, and for each incident energy we employ $2,000$ events.
As the incident energy increases, the reaction becomes more violent, and the apparent temperature of the reaction increases.
In fragmentation reaction, the projectile and target nuclei initially form a compound-like system after they collide each other.
In the early stage of the reaction, only a small number of nucleons evaporate or eject, so the mass of the compound-like system approximately equals to the sum of the projectile and the target.
We exhibit in Fig.~\ref{fig.3} the time evolution of the central density of the heaviest fragment formed in $\isotope[103]{Pd}$ $+$ $\isotope[9]{Be}$ reaction at different incident energies from the IQMD model.
The density in the IQMD model is obtained through the sum of the single-particle density,
\begin{equation}\label{eq.3}
\rho (\vec{r})= \sum^{A}_{i=1}\rho_{i} = \sum^{A}_{i=1}\frac{1}{(2{\pi}L)^{3/2}}{\rm exp}\bigg\{-\frac{\big[\vec{r}-\vec{r}_{i}(t)\big]^{2}}{2L}\bigg\}.
\end{equation}
We notice from the figure that the central density of the heaviest fragment exhibits some oscillations at the beginning. 
This reflects the breathing mode caused by the initial compression of the system, since before the compound-like system dismantling, the largest fragment is the compound-like system itself.
The black dotted line represents $\rho$ $=$ $0.156~\rm fm^{-3}$, which is the initial central density of the nuclear source (\isotope[112]{Sn}) at finite temperature we generate for training the DNN. 
Therefore, the initial compound-like system with density around nuclear saturation density can be mimicked reasonably by an excited nuclear source with certain given temperature.

\begin{figure}[htbp]
\includegraphics
 [width=7.5cm]{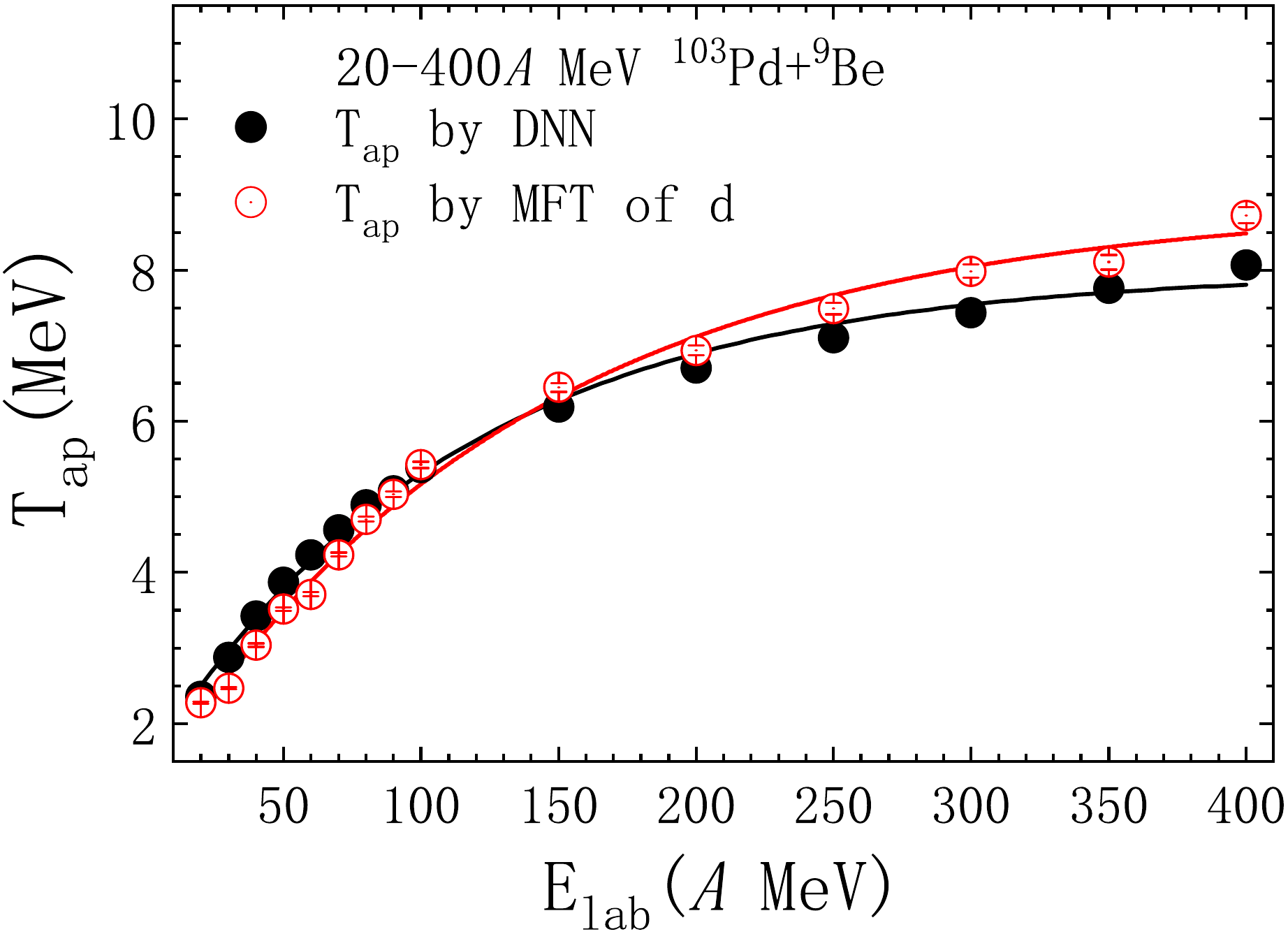}
\caption{The apparent temperature $T_{\rm ap}$ of the reaction $\isotope[103]{Pd}$ $+$ $\isotope[9]{Be}$ at different $E_{\rm lab}$, predicted by DNN (black dots) and MFT (red dots).
The solid lines are their fittings to guide for the eye.
}\label{fig.4}
\end{figure}

With the final-state charge multiplicity distribution $M_{\rm c}(Z_{\rm cf})$ of the reaction $\isotope[103]{Pd}$ $+$ $\isotope[9]{Be}$ simulated with the IQMD, we obtain their apparent temperature through the trained DNN in Eq.~(\ref{E:DNN}).
We plot the obtained apparent temperature $T_{\rm ap}$ of the reaction \isotope[103]{Pd} $+$ \isotope[9]{Be} at different incident energies $E_{\rm lab}$ in Fig.~\ref{fig.4}.
Since the finite temperature nuclear source generated to train the DNN is initialized at around nuclear saturation density, $T_{\rm ap}$ obtained through the present method actually reflects the apparent temperature of the early stage of the compound system in reaction \isotope[103]{Pd} $+$ \isotope[9]{Be}.
In Fig.~\ref{fig.4}, the black symbols are the predicted $T_{\rm ap}$ of compound system at the early stage of the reaction by DNN.
We further test the effects of imperfect acceptance and efficiency on the obtained $T_{\rm ap}$, by applying an acceptance and efficiency cut, respectively, in the IQMD simulations.
It shows that these effects are negligible, which indicates the robustness of the present method on imperfect experimental acceptance and efficiency.

In order to provide a crosscheck of the present method, we employ a momentum fluctuation thermometer (MFT)~\cite{WueNPA843} to determine the apparent temperature of the reaction $\isotope[103]{Pd}$ $+$ $\isotope[9]{Be}$ simulated by IQMD.
In MFT, the distribution of a certain species of light fragment~(deuteron in the present work) is assumed to be Maxwellian, and the temperature of the system is related to the variance of its quadrupole moment $\sigma$ through
\begin{equation}\label{E:MFT}
\sigma^2 = 4A^2m^2_{0}T^2,
\end{equation}
where $m_0$ is the mass of a nucleon and $A$ is the mass number of the fragment.
We add in Fig.~\ref{fig.4} the apparent temperature predicted by MFT with red dots.
We notice that the $T_{\rm ap}$ of the two methods are very close, which increased the reliability of the extracted apparent temperature.

\subsection{Caloric curve}

\begin{figure}[htb]
\includegraphics
[width=7.5cm]{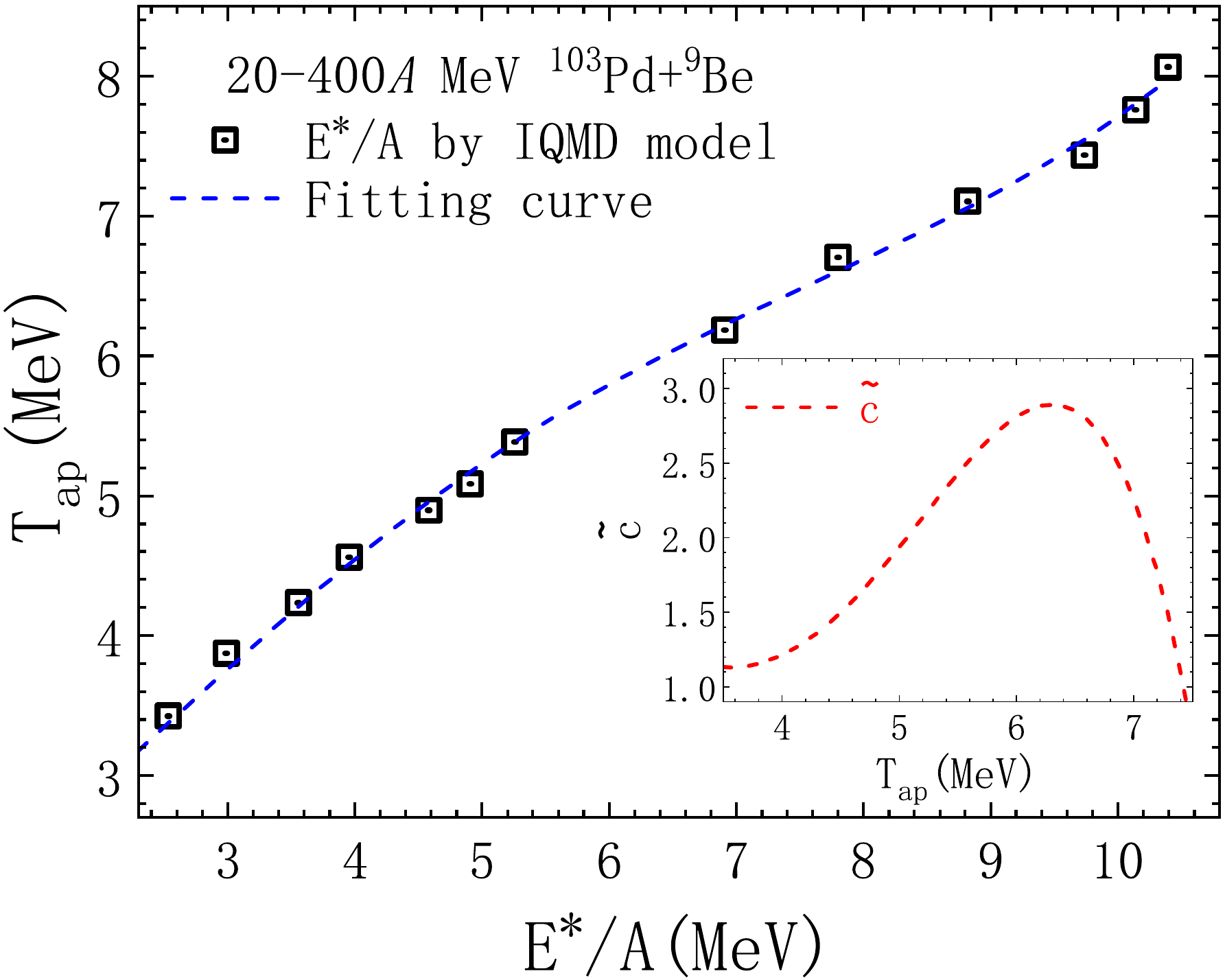}
\caption{Caloric curve of the reaction $\isotope[103]{Pd}$ $+$ $\isotope[9]{Be}$.
Black open squares represent the result based on the IQMD model, with $T_{\rm ap}$ determined by DNN using $M_{\rm c}(Z_{\rm cf})$.
Blue dashed line is its polynomial fit.
The inset shows the specific heat capacity $\tilde{c}$ derived from the fitted formula.}\label{fig.5}
\end{figure}

The caloric curve, i.e., the apparent temperature as a function of excitation energy per nucleon $E^*/A$ of HICs has been considered as an important probe to the existence of the nuclear liquid-gas phase transition \cite{PocPRL75,NatPRC65,NatPRL89,WadPRC99}.
In order to examine the validity of $T_{\rm ap}$ determined by DNN using the charge multiplicity distribution $M_{\rm c}(Z_{\rm cf})$, we study the caloric curve of the reaction $\isotope[103]{Pd}$ $+$ $\isotope[9]{Be}$.
In the IQMD model,  the excitation energy of the compound system at the early stage of the reaction $\isotope[103]{Pd}$ $+$ $\isotope[9]{Be}$ can be obtained by 
\begin{equation}
E^* = U + E_\text{k} - E_\text{0},
\end{equation}
where $U$, $E_\text{k}$ and $E_\text{0}$ are the potential energy, kinetic energy and experimental binding energy \cite{AME2016}, respectively.
To properly account the energy deposited in the system, the kinetic energy of emitted or evaporated nucleons, should be counted when calculating the excitation energy.

We exhibit in Fig.~\ref{fig.5} the caloric curve of the $\isotope[103]{Pd}$ $+$ $\isotope[9]{Be}$ reaction from the IQMD simulation with the apparent temperature determined by DNN using $M_{\rm c}(Z_{\rm cf})$ of the reaction.
As shown in the figure, the increase of $T_{\rm ap}$ slows down when $E^*/A$ reaches to about $8~\rm MeV$.
Traditionally, this characteristic behavior of the caloric curve is explained that, as the excitation energy increases, the system is driven to a spinodal region, in which part of the excitation energy begins to transfer to latent heat.
To characterize this feature of caloric curve quantitatively, the specific heat capacity of the system \cite{ChoPRL85} is defined 
\begin{equation}
\tilde{c}\equiv\frac{d(E^*/A)}{dT_{\rm ap}}.
\end{equation}
Note that it is different from $c_p$ or $c_v$ because it  is not practicable to maintain the external condition on pressure or volume during the reaction.
The apparent temperature corresponding to the maximum of $\tilde{c}$ is called limiting temperature $T_{\rm lim}$, which can be used to deduce the critical temperature of nuclear matter~\cite{NatPRL89}.
We further obtain $\tilde{c}$ through a polynomial fit (red dashed line in Fig.~\ref{fig.5}) of the obtained caloric curve, as shown in the inset of Fig.~\ref{fig.5}.
Based on $T_{\rm ap}$ obtained by DNN from $M_{\rm c}(Z_{\rm cf})$, the obtained $T_{\rm lim}$ of the reaction $\isotope[103]{Pd}$ $+$ $\isotope[9]{Be}$ is about $6.4~\rm MeV$.
This value of $T_{\rm lim}$ follows the general trend of the Natowitz's limiting-temperature dependence to the system size~\cite{NatPRC65}, and thus indicates the validity of determining the apparent temperature through charge multiplicity distribution presented in this article.

\section{Summary and outlook}\label{4}
In the present work, we have examined the possibility of determining the apparent temperature $T_{\rm ap}$ of HICs at intermediate-to-low energies through their final-state charge multiplicity distribution $M_{\rm c}(Z_{\rm cf})$.
Based on the IQMD simulations of de-exciting nuclear sources (\isotope[112]{Sn}) at given temperatures, we have established a relation between the final-state $M_{\rm c}(Z_{\rm cf})$ of a nuclear source, and its corresponding temperature, through training a DNN.
The trained DNN can predict the apparent temperature within an error of $0.62~\rm MeV$, which is small enough for applying it to analyze the reaction dynamics.
We have then employed the above method to obtain the apparent temperature of the $\isotope[103]{Pd}$ $+$ $\isotope[9]{Be}$ reactions at different incident energies simulated by the IQMD, and subsequently the caloric curve.
The caloric curve shows a characteristic behavior at, i.e., $T_{\rm lim}$ $=$ $6.4~\rm MeV$, which follows the general trend of the limiting temperature's dependence to system size and indicates of  nuclear liquid-gas phase transition of a given system.
The present method provides an alternative way to determine the apparent temperature of the HICs at intermediate-to-low energy, and it can be used as a supplement of the traditional nuclear thermometers.
To apply the present method to the analyses of experimental data relies on the accurate description of the fragments multiplicity from dynamical model.
One of the possible ways to achieve this goal is using transport model plus statistical multifragmentation process.
Studies following this line are in progress.

\vspace{0.5cm}
\section*{Declaration of competing interest}
The authors declare that they have no known competing financial interests or personal relationships that could have appeared to influence the work reported in this paper.

\begin{acknowledgments}
This work is partially supported by the National Natural Science Foundation of China under Contracts No. $11890710$, $11890714$ and No. $11625521$, the Key Research Program of Frontier Sciences of the CAS under Grant No. QYZDJ-SSW-SLH$002$, the Strategic Priority Research Program of the CAS under Grants No. XDB34000000,  Guangdong Major Project of Basic and Applied Basic Research No. 2020B0301030008, and the Postdoctoral Innovative Talent Program of China under Grants No. BX$20200098$.
\end{acknowledgments}


\end{CJK*}
\end{document}